\documentclass{llncs}
\usepackage{llncsdoc}

\usepackage{amsmath,amsfonts,amssymb,bbm,subfigure}
\usepackage{algorithm,algorithmic,enumitem,comment}

\usepackage{tikz}
\usetikzlibrary{arrows}

\newtheorem{result}{Result}

\def\EE{{\mathbb{E}}}
\def\PP{{\mathbb{P}}}

\newcommand{\fall}{\,\forall\,}

\newcommand{\PR}{\mathbf{\pi}}

\newcommand{\epr}{r_{\text{max}}}

\newcommand{\abs}[1]{\left \lvert #1 \right \rvert}

\newcommand{\pn}[1]{\left ( #1 \right )}
\newcommand{\bk}[1]{\left [ #1 \right ]}
\newcommand{\pfail}{p_{\text{fail}}}

\newcommand{\pluseq}{\mathrel{+}=}
\def\vecE{\mathbf{\underline{e}}}

\def\vecS{\mathbf{\sigma}}
\def\S{\mathcal{S}}
\newcommand{\dbar}{\overline{d}}

\newcommand{\rev}{\color{red}}

\usepackage[colorlinks,
            linkcolor=blue,
            citecolor=blue,
            urlcolor=magenta,
            linktocpage,
            plainpages=false,
            pdftex]{hyperref}

\begin{document}

\title{Bidirectional PageRank Estimation: From Average-Case to Worst-Case}

\author{Peter Lofgren\inst{1} \and Siddhartha Banerjee\inst{2} \and Ashish Goel\inst{1}}
\institute{Stanford University, Stanford CA 94305, USA
\and
Cornell University, Ithaca NY 14850, USA}

\maketitle
\begin{abstract}
We present a new algorithm for estimating the Personalized PageRank (PPR) between a source and target node on undirected graphs, with sublinear running-time guarantees over the worst-case choice of source and target nodes.
{\rev }
Our work builds on a recent line of work on bidirectional estimators for PPR, which obtained sublinear running-time guarantees but in an average-case sense, for a uniformly random choice of target node.
Crucially, we show how the reversibility of random walks on undirected networks can be exploited to convert average-case to worst-case guarantees.  
While past bidirectional methods combine forward random walks with reverse local pushes, our algorithm combines forward local pushes with reverse random walks.
We also discuss how to modify our methods to estimate random-walk probabilities for any length distribution, thereby obtaining fast algorithms for estimating general graph diffusions, including the heat kernel, on undirected networks.
%Whether such worst-case running-time results extend to general graphs, or if PageRank computation is fundamentally easier on undirected graphs as opposed to directed graphs, remains an open question.
\end{abstract}

%!TEX root = main.tex
\section{Introduction}

Ever since their introduction in the seminal work of Page et al. \cite{Page1999}, PageRank and Personalized PageRank (PPR) have become some of the most important and widely used network centrality metrics (a recent survey \cite{Gleich-preprint-pagerank-beyond} lists several examples). 
At a high level, for any graph $G$, given `teleport' probability $\alpha$ and a `personalization distribution' $\vecS$ over the nodes of $G$, PPR models the importance of every node from the point of view of $\vecS$ in terms of the stationary probabilities of `short' random walks that periodically restart from $\vecS$ with probability $\alpha$.
It can be defined recursively as giving  importance $\alpha$ to $\vecS$, and in addition giving every node importance based on the importance of its in-neighbors.

Formally, given normalized adjacency matrix $W=D^{-1}A$, the Personalized PageRank vector $\PR_{\vecS}$ with respect to source distribution $\vecS$ is the solution to
\begin{align}
\label{eq:pr1}
\PR_{\vecS} = \alpha\vecS + (1-\alpha)\PR_{\vecS}W.
\end{align}
An equivalent definition is in terms of the terminal node of a random-walk starting from $\sigma$. 
Let $\{X_0,X_1,X_2,\ldots\}$ be a random-walk starting from $X_0\sim\sigma$, and $L\sim Geometric(\alpha)$. 
Then the PPR of any node $t$ is given by~\cite{Avrachenkov2007}:
\begin{align}
\label{eq:pr2}
\pi_\sigma(t) = \PP[X_L = t]
\end{align}
The equivalence of these definitions can be seen using a power series expansion.

In this work, we focus on developing PPR-estimators with \emph{worst-case sublinear guarantees} for \emph{undirected graphs}. 
Apart from their technical importance, our results are are of practical relevance as several large-scale applications of PPR are based on undirected networks.
For example, Facebook (which is an undirected social network) used Personalized PageRank for friend recommendation \cite{backstrom2011supervised}.  
The social network Twitter is directed, but Twitter's friend recommendation algorithm (Who to Follow) \cite{gupta2013wtf} uses an algorithm called personalized SALSA \cite{lempel2000stochastic,bahmani2010},
%, which merges PageRank with the notion of hubs and authorities introduced in the earlier HITS algorithm \cite{HITS}.
which first converts the directed network into an expanded undirected graph~\footnote{Specifically, for each node $u$ in the original graph, SALSA creates two virtual nodes, a ``consumer-node'' $u'$ and a ``producer-node'' $u''$, which are linked by an undirected edge. Any directed edge $(u, v)$ is then converted into an undirected edge $(u', v'')$ from $u$'s consumer node to $v$'s producer node.}, and then computes PPR on this new graph. 
Random walks have also been used for collaborative filtering by the YouTube team \cite{baluja2008video} (on the undirected user-item bipartite graph), to predict future items a user will view.  
Applications like this motivate fast algorithms for PPR estimation on undirected graphs.

Equations \eqref{eq:pr1} and \eqref{eq:pr2} suggest two natural estimation algorithms for PPR -- via linear-algebraic iterative techniques, and using Monte Carlo. The linear algebraic characterization of PageRank in Eqn. \eqref{eq:pr1} suggests the use of power iteration (or other localized iterations; cf Section \ref{ssec:relwork} for details), while Eqn. \eqref{eq:pr2} is the basis for a Monte-Carlo algorithm, wherein we estimate $\PR_{\vecS}[t]$ by sampling independent $L$-step paths, each starting from a random state sampled from $\vecS$.  
For studying PageRank estimation algorithms, smaller probabilities are more difficult to estimate than large ones, so a natural parametrization is in terms of the minimum PageRank we want to detect.  Formally, given any source $\vecS$, target node $t\in V$ and a desired minimum probability threshold $\delta$, we want algorithms that give accurate estimates whenever $\PR_{\vecS}[t] \geq \delta$.  
Improved algorithms are motivated by the slow convergence of these algorithms: \emph{both Monte Carlo and linear algebraic techniques have a running time of $\Omega(1/\delta)$} for PageRank estimation. Furthermore this is true not only for worst case choices of target state $t$, but on average Monte-Carlo requires $\Omega(1/\delta)$ time to estimate a probability of size $\delta$.  Power iteration takes $\Theta(m)$ time, where $m$ is the number of edges, and the work \cite{Lofgren2013} shows empirically that the local version of power-iteration scales with $1/\delta$ for $\delta > 1/m$.  
 %rather, for many Markov chains %(in particular, Markov chains with high conductance \Peter{(What is the conductance of a Markov Chain?)}, this is true for most choices of target states (cf. Section \ref{ssec:relwork}).  

In a recent line of work, linear-algebraic and Monte-Carlo techniques were combined to develop new \emph{bidirectional PageRank} estimators \texttt{FAST-PPR}~\cite{fastppr} and \texttt{Bidirectional-PPR}~\cite{bidir-search}, which gave the first significant improvement in the running-time of PageRank estimation since the development of Monte-Carlo techniques. Given an arbitrary source distribution $\sigma$ and a \emph{uniform random target node} $t$, these estimators were shown to return an accurate PageRank estimate with an \emph{average} running-time of $\tilde{O} \pn{\sqrt{\overline{d}/\delta}}$, where $\overline{d}=m/n$ is the average degree of the graph.  Given $\tilde{O} \pn{n \sqrt{\overline{d}/\delta}}$ precomputation and storage, the authors prove worst case guarantees for this bidirectional estimator but in practice that is a large precomputation requirement.  This raised the challenge of designing an algorithm with similar running-time guarantees over a \emph{worst-case} choice of target node $t$.  Inspired by the bidirectional estimators in \cite{fastppr,bidir-search}, we propose a new PageRank estimator for \emph{undirected graphs} with \emph{worst-case} running time guarantees. 

\subsection{Our Contribution}
\label{ssec:results}
%While past bidirectional estimators \cite{fastppr,bidir-search} use linear algebra working from the target and sample random walks forwards from the source, we present a novel estimator which samples random walks from the target and uses linear algebra from the source node.  
We present the first estimator for personalized PageRank with sublinear running time in the worst case on undirected graphs.  We formally present our \texttt{Undirected-BiPPR} algorithm in Section \ref{sec:algo}, and prove that it has the following accuracy and running-time guarantees:
\begin{result}[See Theorem \ref{thm:accuracy} in Section \ref{sec:algo}]
Given any undirected graph $G$, teleport probability $\alpha$, source node $s$, target node $t$, threshold $\delta$ and relative error $\epsilon$, the \texttt{Undirected-BiPPR} estimator (Algorithm \ref{alg:BIPPR}) returns an unbiased estimate $\widehat{\PR}_{s}[t]$ for $\PR_{s}[t]$, which, with probability greater than $1-p_{fail}$, satisfies:
\begin{align*}
\left|\widehat{\PR}_{s}[t]-\PR_{s}[t]\right|<\max\left\{\epsilon\PR_{s}[t],2e\delta\right\}.
\end{align*}
\end{result}
%This result follows from straightforward concentration bounds, via arguments which are similar to those for \texttt{Bidirectional-PPR}~\cite{bidir-search}. The novel contribution of this work is the following \emph{parametrized} running-time guarantee:
\begin{result}[See Theorem \ref{thm:worstcase} in Section \ref{sec:algo}]
Let any undirected graph $G$, teleport probability $\alpha$, threshold $\delta$ and desired relative error $\epsilon$ be given.  For any source, target pair $(s,t)$, the \texttt{Undirected-BiPPR} algorithm has a running-time of $O\left(\frac{\sqrt{\ln(1/p_{fail})}}{\epsilon}\sqrt{\frac{d_t}{\delta}}\right)$, where $d_t$ is the degree of the target node $t$.
\end{result}
In personalization applications, we are often only interested in personalized importance scores if they are greater than global importance scores, so it is natural to set $\delta$ based on the global importance of $t$.
Assuming $G$ is connected, in the limit $\alpha\rightarrow 0$, the PPR vector for any start node $s$ converges to the stationary distribution of infinite-length random-walks on $G$ -- that is $\lim_{\alpha\rightarrow 0}\PR_s[t]=d_t/m$. This suggests that a natural PPR significance-test is to check whether $\PR_s(t) \geq d_t/m$. To this end, we have the following corollary:
\begin{result}[See Corollary \ref{corr:worstcase} in Section \ref{sec:algo}]
 For any graph $G$ and any $(s,t)$ pair such that $\pi_s(t) \geq \frac{d_t}{m}$, then with high probability~\footnote{Following convention, we use w.h.p. to mean with probability greater than $1-\frac{1}{n}$.}, \texttt{Undirected-BiPPR} returns an estimate $\pi_s(t)$ with relative error $\epsilon$ with a \emph{worst-case} running-time of $O\pn{\sqrt{m}\log n/\epsilon}$. 
\end{result}	 

Finally, in Section \ref{sec:extension}, using ideas from~\cite{fastmstp}, we extend our technique to estimating more general random-walk transition-probabilities on undirected graphs, including graph diffusions and the heat kernel~\cite{Chung2007,Kloster2014}.

\subsection{Existing Approaches for PageRank Estimation}
\label{ssec:relwork}

We first summarize the existing methods for PageRank estimation:

\noindent\textbf{Monte Carlo Methods}: A standard method~\cite{Avrachenkov2007,borgs2012} for estimating $\PR_{\vecS}[t]$ is by using the terminal node of independently generated random walks of length $L\sim Geometric(\alpha)$ starting from a random node sampled from $\vecS$. Simple concentration arguments show that we need $\widetilde{\Theta}(1/\delta)$ samples to get an accurate estimate of $\PR_{\vecS}[t]$, irrespective of the choice of $t$ and graph $G$.

\noindent\textbf{Linear-Algebraic Iterations}: Since the PageRank vector is the stationary distribution of a Markov chain, it can also be estimated via forward or reverse power iterations. A direct power iteration is often infeasible for large graphs; in such cases, it is preferable to use localized power iterations~\cite{Andersen2006,Andersen2007}. These local-update methods can also be used for other transition probability estimation problems such as heat kernel estimation \cite{Kloster2014}. Local update algorithms are often fast in practice, as unlike full power iteration methods they exploit the local structure of the chain. However even in sparse Markov chains and for a large fraction of target states, their running time can be $\Omega(1/\delta)$. 
For example, consider a random walk on a random $d$-regular graph and let $\delta= o(1/n)$.  Then for $\ell\sim\log_d(1/\delta)$, verifying $\PR_{\vecE_s}[t]>\delta$ is equivalent to uncovering the entire $\log_d(1/\delta)$ neighborhood of $s$. However since a large random $d$-regular graph is (w.h.p.) an expander, this neighborhood has $\Omega(1/\delta)$ distinct nodes.

\noindent\textbf{Bidirectional Techniques}: Bidirectional methods are based on simultaneously working forward from the source node $s$ and backward from the target node $t$ in order to improve the running-time. One example of such a bidirectional technique is the use of \emph{colliding random-walks} to estimate length-$2\ell$ random-walk transition probabilities in \emph{regular undirected graphs}~\cite{Goldreich2011,Kale2008} -- the main idea here is to exploit the reversibility by using two independent random walks of length $\ell$ starting from $s$ and $t$ respectively, and detecting if they collide. This results in reducing the number of walks required by a square-root factor, based on an argument similar to the birthday-paradox.

The \texttt{FAST-PPR} algorithm of Lofgren et al. \cite{fastppr} was the first bidirectional algorithm for estimating PPR in general graphs; this was subsequently refined and improved by the \texttt{Bidirectional-PPR} algorithm \cite{bidir-search}, and also generalized to other Markov chain estimation problems~\cite{fastmstp}. 
These algorithms are based on using a reverse local-update iteration from the target $t$ (adapted from Andersen et al.~\cite{Andersen2007}) to smear the mass over a larger \emph{target set}, and then using random-walks from the source $s$ to detect this target set.
From a theoretical perspective, a significant breakthrough was in showing that for arbitrary choice of source node $s$ these bidirectional algorithms achieved an \emph{average} running-time of $\tilde{O}(\sqrt{\overline{d}/\delta})$ over uniform-random choice of target node $t$ -- in contrast, both local-update and Monte Carlo has a running-time of $\Omega(1/\delta)$ for uniform-random targets. 
More recently, \cite{bressan2014} showed that a similar bidirectional technique achieved a sublinear query-complexity for global PageRank computation, under a modified query model, in which all neighbors of a given node could be found in $O(1)$ time.

%!TEX root = main.tex

\section{PageRank Estimation in Undirected Graphs}
\label{sec:algo}

We now present our new bidirectional algorithm for PageRank estimation in undirected graphs.

\subsection{Preliminaries}

We consider an undirected graph $G(V,E)$, with $n$ nodes and $m$ edges.
For ease of notation, we henceforth consider unweighted graphs, and focus on the simple case where $\sigma = \vecE_s$ for some single node $s$. 
We note however that all our results extend to weighted graphs and any source distribution $\sigma$ in a straightforward manner.% (with some additional notation and assumptions).

%Recall that in a weighted directed graph, the normalized adjacency matrix obeys $W=D^{-1}A$, where $A$ is the matrix of edge weights $w_{ij}$, and $D = diag(d_i)$ is a diagonal-matrix with $d_i:=\sum_jw_{ij}$. 
%In an un-weighted graph (i.e., if all edge-weights are $1$), then $A$ is the adjacency matrix and $d_i$ the out-degree of node $i$. The case where $A$ is symmetric corresponds to a (weighted) undirected graph. 

\subsection{A Symmetry for PPR in Undirected Graphs}

%Consider an undirected graph $G(V,E)$, with $n$ nodes and $m$ edges. The edges have associated edge-weights $w_{ij}\fall(i,j)\in E$ (and $w_{ij}=w_{ji}$); recall that we define $d_i=\sum_j w_{ij}\fall i\in V$. The probability that a random-walk transitions from node $i$ to node $j$ is then given by $(D^{-1}A)_{ij} = w_{ij}/d_i$, and from $j$ to $i$ by $w_{ij}/d_j$. 

The \texttt{Undirected-BiPPR} Algorithm critically depends on an underlying \emph{reversibility property} exhibited by PPR vectors in undirected graphs. 
This property, stated before in several earlier works \cite{avrachenkov2013,Grolmusz2015}, is a direct consequence of the reversibility of random walks on undirected graphs.
To keep our presentation self-contained, we present this property, along with a simple probabilistic proof, in the form of the following lemma:
\begin{lemma} 
\label{undirected_ppr_lemma} 
Given any undirected graph $G$, for any teleport probability $\alpha \in (0,1)$ and for any node-pair $(s, t) \in V^2$, we have:
\[ \pi_s[t] =  \frac{d_t}{d_s} \pi_t[s]. \]  
\end{lemma}

\begin{proof}
For path $P=\{s,v_1,v_2,\ldots,v_k,t\}$ in $G$, we denote its length as $\ell(P)$ (here $\ell(P) = k+1$), and define its reverse path to be $\overline{P}=\{t,v_k,\ldots,v_2,v_1,s\}$ -- note that $\ell(P) = \ell(\overline{P})$. 
Moreover, we know that a random-walk starting from $s$ traverses path $P$ with probability 
$\PP[P]= \frac{1}{d_s}\cdot \frac{1}{d_{v_1}}\cdot \ldots \cdot \frac{1}{d_{v_k}}$,
and thus, it is easy to see that we have:
\begin{equation}
\label{eq:path}
\PP[P]\cdot d_s = \PP[\overline{P}]\cdot d_t 
\end{equation}
Now let $\mathcal{P}_{st}$ denote the set of paths in $G$ starting at $s$ and terminating at $t$. Then we can re-write Eqn. \eqref{eq:pr2} as:
\begin{align*}
	\PR_s[t] = \prod_{P\in \mathcal{P}_{st}}\alpha(1-\alpha)^{\ell(P)}\PP[P] 
	= \prod_{\overline{P}\in \mathcal{P}_{ts}}\alpha(1-\alpha)^{\ell(\overline{P})}\PP[\overline{P}]
	= \frac{d_t}{d_s} \PR_t[s]\quad\square
\end{align*}
\end{proof}

%\begin{proof}[of Lemma \ref{undirected_ppr_lemma}]
%We simply use the random walk interpretation of PageRank to write down the probability of a walk occurring as a sum over the set of all $s$-$t$ paths of a given length $l$.  Let $P_l = \{(p_0, p_1, \ldots, p_l) \in V^{l+1}: p_0=s, p_l=t, \forall i \in \{0,\ldots, l-1\} (p_i, p_{i+1}) \in E\}$. Then
%\begin{align*}
% \pi_v[t] &= \sum_{l=0}^{\infty} (1-\alpha)^l \alpha \sum_{p \in P_l} \prod_{i=0}^{l-1} \frac{1}{d_{p_i}}  \\
% &= \frac{d_t}{d_v} \sum_{l=0}^{\infty} (1-\alpha)^l \alpha \sum_{p \in P_l} \prod_{i=1}^{l} \frac{1}{d_{p_i}}  \\
%&=  \frac{d_t}{d_v} \pi_t[v].   \quad\square
%\end{align*}
%\end{proof}

\subsection{The \texttt{Undirected-BiPPR} Algorithm} 

At a high level, the \texttt{Undirected-BiPPR} algorithm has two components:
\begin{itemize}[nosep,leftmargin=*]
\item \textbf{Forward-work}: Starting from source $s$, we first use a forward local-update algorithm, the \texttt{ApproximatePageRank}$(G,\alpha,s,\epr)$ algorithm of Andersen et al.~\cite{Andersen2006} (shown here as Algorithm \ref{alg:forwardPush}). This procedure begins by placing one unit of ``residual'' probability-mass on $s$, then repeatedly selecting some node $u$, converting an $\alpha$-fraction of the residual mass at $u$ into probability mass, and pushing the remaining residual mass to $u$'s neighbors.  For any node $u$, it returns an estimate $p_s[u]$ of its PPR $\PR_s[u]$ from $s$ %with additive error at most $d_u\epr$.
 as well as a residual $r_s[u]$ which represents un-pushed mass at $u$.
\item \textbf{Reverse-work}:  
We next sample random walks of length $L\sim Geometric(\alpha)$ starting from $t$, and use the residual at the terminal nodes of these walks to compute our desired PPR estimate. Our use of random walks backwards from $t$ depends critically on the symmetry in undirected graphs presented in Lemma \ref{undirected_ppr_lemma}.
\end{itemize}
Note that this is in contrast to \texttt{FAST-PPR} and \texttt{Bidirectional-PPR}, which performs the local-update step in reverse from the target $t$, and generates random-walks forwards from the source $s$.

\begin{algorithm}[!ht]
\caption{\texttt{ApproximatePageRank}$(G,\alpha, s,\epr)$~\cite{Andersen2006}}
\label{alg:forwardPush}
\begin{algorithmic}[1]
\REQUIRE graph $G$, teleport probability $\alpha$, start node $s$, maximum residual $\epr$
\STATE Initialize (sparse) estimate-vector $p_s = \vec{0}$ and (sparse) residual-vector $r_s = e_s$ (i.e.~$r_s[v] = 1$ if $v=s$; else $0$)

\WHILE{$\exists u\in V\,s.t.\, \frac{r_s[u]}{d_u}> \epr$}
\FOR{$v\in \mathcal{N}[u]$}
   \STATE $r_s[v] \pluseq (1 - \alpha) r_s[u] / d_u$
\ENDFOR
\STATE      $p_s[u] \pluseq \alpha r_s[u]$
\STATE      $r_s[u]=0$
\ENDWHILE
\RETURN $(p_s, r_s)$
\end{algorithmic}
\end{algorithm}

In more detail, %our forward-work component is based on the local update algorithm of \cite{Andersen2006} (shown here as Algorithm \ref{alg:forwardPush}) which works forwards from a start $s$ to compute estimates $p_s \in \R^n$ of $\pi_s$ and residuals $r_s \in \R^n$ which represent mass not yet pushed forward.
our algorithm will choose a maximum residual parameter $\epr$, and apply the local push operation in Algorithm \ref{alg:forwardPush} until for all $v$, $r_s[v]/d_v < \epr$. Andersen et al.~\cite{Andersen2006} prove that their local-push operation preserves the following invariant for vectors $(p_s,r_s)$:
\begin{equation}
  \label{eq:forward_push}
 \pi_s[t] = p_s[t] + \sum_{v \in V} r_s[v] \pi_v[t],\qquad\fall t\in V .  
\end{equation}
Since we ensure that $\forall v, r_s[v]/d_v < \epr$, it is natural at this point to use the symmetry Lemma \ref{undirected_ppr_lemma} and re-write this as:
\[ \pi_s[t] = p_s[t] + d_t \sum_{v \in V}  \frac{r_s[v]}{d_v} \pi_t[v]. \]
Now using the fact that $\sum_t \pi_v[t]=n\pi[t]$ get that $\fall t\in V$, $ \left|\pi_s[t] - p_s[t]\right|  \leq \epr d_t n \pi[t]$.

However, we can get a more accurate estimate by using the residuals.  
The key idea of our algorithm is to re-interpret this as an expectation:
\begin{align}
\label{eq:estimate}
\pi_s[t] = p_s[t] + d_t  \EE_{V \sim \pi_t} \bk{\frac{r_s[v]}{d_V} }.
\end{align}
We estimate the expectation using standard Monte-Carlo.  Let $V_i \sim \pi_t$ and $X_i = r_s(V_i) d_t/d_{V_i}$, so we have $\pi_s[t] = p_s[t] + \EE[X]$. Moreover, each sample $X_i$ is bounded by $d_t \epr$ (this is the stopping condition for \texttt{ApproximatePageRank}), which allows us to efficiently estimate its expectation.
To this end, we generate $w$ random walks, where
\[ w = \frac{c}{\epsilon^2} \frac{\epr}{\delta / d_t}.
 \]
The choice of $c$ is specified in Theorem \ref{thm:accuracy}. Finally, we return the estimate:
\[ \widehat{\pi}_s[t] = p_t[s] + \frac{1}{w} \sum_{i=1}^w X_i. \]
The complete pseudocode is given in Algorithm \ref{alg:BIPPR}.

\begin{algorithm}[!ht]
\caption{\texttt{Undirected-BiPPR}$(s,t,\delta)$}
\label{alg:BIPPR}
\begin{algorithmic}[1] 
\REQUIRE graph $G$, teleport probability $\alpha$, start node $s$, target node $t$, minimum probability $\delta$, accuracy parameter $c= 3\ln \pn{2/\pfail}$ \quad(cf. Theorem \ref{thm:accuracy})

\STATE $(p_s, r_s)$ =  \texttt{ApproximatePageRank}$(s,\epr)$

\STATE Set number of walks $w=c d_t \epr/(\epsilon^2\delta)$ 
\FOR{index $i\in [w]$}
\STATE Sample a random walk starting from $t$, stopping after each step with probability $\alpha$; let $V_i$ be the endpoint
\STATE Set $X_i = r_s(V_i) / d_{V_i}$
\ENDFOR 
\RETURN $\widehat{\PR}_s[t]=p_s[t] + (1/w)\sum_{i\in[w]}X_i$
\end{algorithmic}
\end{algorithm}

\subsection{Analyzing the Performance of \texttt{Undirected-BiPPR}}

\noindent\textbf{Accuracy Analysis}: We first prove that \texttt{Undirected-BiPPR} returns an unbiased estimate with the desired accuracy:

\begin{theorem}
\label{thm:accuracy} 
In an undirected graph $G$, for any source node $s$, minimum threshold $\delta$, maximum residual $\epr$, relative error $\epsilon$, and failure probability $\pfail$, Algorithm \ref{alg:BIPPR} outputs an estimate $\widehat{\PR}_s[t]$ such
that with probability at least $1- \pfail$ we have: $\qquad\abs{\pi_s[t]-\hat{\pi}_s[t]} \leq \max\{\epsilon\pi_s[t],2e\delta\}$. 

%\begin{itemize}[nosep,leftmargin=*]
%\item If $\pi_s[t] \geq \delta$: \hspace{1cm} $\abs{\pi_s[t]-\hat{\pi}_s[t]} \leq \epsilon \pi_s[t]$. 
%\item If $\pi_s[t] \leq \delta$: \hspace{1cm}
%$\abs{\pi_s[t]-\hat{\pi}_s[t]} \leq 2e\delta$.
%\end{itemize}
\end{theorem}

%The above result shows that the estimate $\hat{\pi}_s[t]$ can be used to distinguish between `significant' and `insignificant' PPR pairs: for pair $(s,t)$, Theorem \ref{thm:accuracy} guarantees that if $\pi_s[t] \geq \frac{(1+2e)\delta}{(1-\epsilon)}$, then the estimate is greater than $(1+2e)\delta$, whereas if $\pi_s[t] < \delta$, then the estimate is less than $(1+2e)\delta$. 

The proof follows a similar outline as the proof of Theorem $1$ in \cite{bidir-search}. For completeness, we sketch the proof here:

\begin{proof} 
As stated in Algorithm \ref{alg:BIPPR}, we average over $w= cd_t\epr/\epsilon^2\delta$ walks, where $c$ is a parameter we choose later.
Each walk is of length $Geometric(\alpha)$, and we denote $V_i$ as the last node visited by the $i^{th}$ walk; note that $V_i\sim\pi_t$.
As defined above, let $X_i = r_s(V_i) d_t/d_{V_i}$; the estimate returned by \texttt{Undirected-BiPPR} is:
\[\widehat{\pi}_s[t] = p_t[s] + \frac{1}{w} \sum_{i=1}^w X_i. \]
First, from Eqn. \eqref{eq:estimate}, we have that $\EE[\widehat{\pi}_s[t]]= \pi_s[t]$.   
Also, \texttt{ApproximatePageRank} guarantees that for all $v$, $r_s[v]<d_v\epr$, and so each $X_i$ is bounded in $[0,d_t\epr]$; for convenience, we rescale $X_i$ by defining $Y_i = \frac{1}{d_t\epr} X_i$.

We now show concentration of the estimates via the following Chernoff bounds (see Theorem $1.1$ in~\cite{DuPa09}):
\begin{enumerate}[nosep]
\item $\PP[|Y - \EE[Y]| > \epsilon \EE[Y]] < 2 \exp(-\frac{\epsilon^2}{3}\EE[Y])$
\item $\textrm{For any } b > 2e\EE[Y], \PP[Y > b] \leq 2^{-b}$
\end{enumerate}
We perform a case analysis based on whether $\EE[X_i] \geq \delta$ or $\EE[X_i] < \delta$. 
First, if $\EE[X_i] \geq \delta$, then we have $\EE[Y] = \frac{w}{d_t\epr} \EE[X_i] = \frac{c}{\epsilon^2\delta} \EE[X_i] \geq \frac{c}{\epsilon^2}$, and thus:
\begin{align*}
\PP\left[\abs{\widehat{\pi}_s[t] - \pi_s[t]} > \epsilon \pi_s[t]\right] 
 &\leq \PP\left[\abs{\bar{X} - \EE[X_i]} > \epsilon \EE[X_i]\right]
  = \PP\left[\abs{Y - \EE[Y]} > \epsilon \EE[Y]\right] \\
  &\leq 2 \exp\pn{-\frac{\epsilon^2}{3}\EE[Y]} 
  \leq 2 \exp\pn{-\frac{c}{3} }
  \leq \pfail,
\end{align*}
where the last line holds as long as we choose  $c \geq 3 \ln \pn{2/\pfail}$.
%This case is concluded because if $\abs{\bar{X} - \EE[X_i]} \leq \epsilon \EE[X_i]$ then $\abs{\widehat{\pi_s[t]} - \pi_s_t} \leq \epsilon \pi_s_t$.

Suppose alternatively that $\EE[X_i] < \delta$.  Then:
\begin{align*}
\PP[\abs{\hat{\pi}_s[t] - \pi_s[t]} > 2e\delta]
&= \PP[\abs{\bar{X} - \EE[X_i]} > 2e\delta] 
%&=  \PP\bk{\abs{\frac{\epr}{w} Y - \frac{\epr}{w} \EE[Y]} > 2e\delta} \\
=  \PP\bk{\abs{Y -  \EE[Y]} > \frac{w}{d_t\epr} 2e\delta} \\
&\leq  \PP\bk{Y > \frac{w}{d_t \epr}2e\delta} .
\end{align*}
At this point we set $b = 2e\delta w/d_t\epr=2ec/\epsilon^2$ and apply the second Chernoff bound.  
Note that $\EE[Y] = c\EE[X_i]/\epsilon^2\delta  < c/\epsilon^2$, and hence we satisfy $b > 2e \EE[Y]$.
We conclude that:
\[ \PP[\abs{\hat{\pi}_s[t] - \pi_s[t]} > 2e\delta] \leq 2^{-b} \leq \pfail \]
as long as we choose $c$ such that $c \geq \frac{\epsilon^2}{2e} \log_2 \frac{1}{\pfail}$. The proof is completed by combining both cases and choosing $c = 3 \ln \pn{2/\pfail}$.$\quad\square$
\end{proof}

\noindent\textbf{Running Time Analysis}: The more interesting analysis is that of the running-time of \texttt{Undirected-BiPPR} -- we now prove a worst-case running-time bound:

\begin{theorem}
\label{thm:worstcase}
  In an undirected graph, for any source node (or distribution) $s$, target $t$ with degree $d_t$, threshold $\delta$, maximum residual $\epr$, relative error $\epsilon$, and failure probability $\pfail$, \texttt{Undirected-BiPPR} has a worst-case running-time of:
\[ O \pn{ \frac{\sqrt{\log{\frac{1}{\pfail}}}}{\epsilon} \sqrt{\frac{d_t}{\delta}}}. \]
\end{theorem}

Before proving this result, we first state and prove a crucial lemma from \cite{Andersen2006}:
\begin{lemma}[Lemma $2$ in \cite{Andersen2006}]
	\label{lem:andersen}
Let $T$ be the total number of push operations performed by \texttt{ApproximatePageRank}, and let $d_k$ be the degree of the vertex involved in the $k^{th}$ push. Then:
\begin{align*}
\sum_{k=1}^Td_k\leq\frac{1}{\alpha\epr}	
\end{align*}
\end{lemma}	
\begin{proof}
Let $v_k$ be the vertex pushed in the $k^{th}$ step -- then by definition, we have that $r_s(v_k)>\epr d_k$. Now after the local-push operation, the sum residual $||r_s||_1$ decreases by at least $\alpha\epr d_k$. However, we started with $||r_s||_1 = 1$, and thus we have $\sum_{k=1}^T\alpha\epr d_k\leq 1$. $\qquad\square$
\end{proof}
Note also that the amount of work done while pushing from a node $v$ is $d_v$.
\begin{proof}[of Theorem \ref{thm:worstcase}]
  As proven in Lemma \ref{lem:andersen}, the push forward step takes total time $O\pn{1/\alpha \epr}$ in the \emph{worst-case}.  The random walks take $O(w) = O\pn{\frac{1}{\epsilon^2} \frac{\epr}{\delta / d_t}}$ time.  Thus our total time is
\[ O\pn{\frac{1}{\epr} + \frac{\ln{\frac{1}{\pfail}}}{\epsilon^2} \frac{\epr}{\delta / d_t}}. \]
Balancing this by choosing $r_{\max}=\frac{\epsilon}{\sqrt{\ln{\frac{1}{\pfail}}}} \sqrt{\delta/d_t}$, we get total running-time:
\[ O \pn{ \frac{\sqrt{\ln{\frac{1}{\pfail}}}}{\epsilon} \sqrt{\frac{d_t}{\delta}}}. \qquad\square\]
\end{proof}
We can get a cleaner worst-case running time bound if we make a natural assumption on $\pi_s[t]$.  In an undirected graph, if we let $\alpha = 0$ and take infinitely long walks, the stationary probability of being at any node $t$ is $\frac{d_t}{m}$.  Thus if $\pi_s[t] < \frac{d_t}{m}$, then $s$ actually has a lower PPR to $t$ than the non-personalized stationary probability of $t$, so it is natural to say $t$ is not significant for $s$.  If we set a significance threshold of $\delta = \frac{d_t}{m}$, and apply the previous theorem, we immediately get the following:
\begin{corollary}
	\label{corr:worstcase}
  If $\pi_s[t] \geq \frac{d_t}{m}$, we can estimate $\pi_s[t]$ within relative error $\epsilon$ with probability greater than $1-\frac{1}{n}$ in worst-case time:
\[ O\pn{\frac{\log n}{\epsilon} \sqrt{m}}. \]
\end{corollary}

In contrast, the running time for \texttt{Monte-Carlo} to achieve the same accuracy guarantee is $O\pn{\frac{1}{\delta} \frac{\log\pn{1/\pfail}}{\alpha \epsilon^2}}$, and the running time for \texttt{ApproximatePageRank} is $O\pn{\frac{\bar{d}}{\delta \alpha}}$.  The \texttt{FAST-PPR} algorithm of \cite{fastppr} has an \emph{average case} running time of $O\pn{\frac{1}{\alpha\epsilon^2}\sqrt{\frac{\bar{d}}{\delta}} \sqrt{ \frac{\log\pn{1/\pfail} \log\pn{1/\delta}}{\log\pn{1/(1-\alpha)}} }}$ for uniformly chosen targets, but has no clean worst-case running time bound because its running time depends on the degree of nodes pushed from in the linear-algebraic part of the algorithm.

%The results are shown in Figure \ref{fig:ppr_runtime}.  
%\begin{figure*}[t]
%\centering
%\subfigure[Sampling targets uniformly]{
%\label{fig:runtime_uniform}
%\includegraphics[width=0.9\columnwidth]{ppr_runtime_uniform.png}
%}
%\hfill
%\subfigure[Sampling targets from PageRank distribution]{
%\label{fig:runtime_pagerank}
%\includegraphics[width=1.1\columnwidth]{ppr_runtime_pagerank.png}
%}
%%\subfigure[Other stuff]{
%%\includegraphics[width=.65\textwidth]{somethingelse.png}}
%\caption[Running-time Plots]{Average running-time (on log-scale) for different networks. We measure the time required for estimating PPR values $\pi_s[t]$ with threshold $\delta=\frac{4}{n}$ for $1000$ $(s,t)$ pairs. For each pair, the start node is sampled uniformly, while the target node is sampled uniformly in Figure \ref{fig:runtime_uniform}, or from the global PageRank distribution in Figure \ref{fig:runtime_pagerank}. In this plot we use teleport probability $\alpha=0.2$.}
%\label{fig:ppr_runtime}
%\end{figure*}

%!TEX root = main.tex

\section{Extension to Graph Diffusions}
\label{sec:extension}

PageRank and Personalized PageRank are a special case of a more general set of network-centrality metrics referred to as \emph{graph diffusions}~\cite{Chung2007,Kloster2014}. 
In a a graph diffusion we assign a weight $\alpha_i$ to walks of length $i$. The score is then is a polynomial function of the random-walk transition probabilities of the form:
\begin{align*}
f(W,\vecS):=\sum_{i=0}^{\infty}\alpha_i\left(\vecS W^{i}\right),
\end{align*}
where $\alpha_i\geq 0, \sum_i\alpha_i = 1$. To see that PageRank has this form, we can expand Eqn.~\eqref{eq:pr1} via a Taylor series to get:
\begin{align*}
\PR_{\vecS} = \sum_{i=1}^{\infty}\alpha(1-\alpha)^i \left(\vecS W^i\right)	
\end{align*}
Another important graph diffusion is the \emph{heat kernel} $h_{\sigma}$, which corresponds to the scaled matrix exponent of $(I-W)^{-1}$:
\begin{align*}
h_{\vecS,\gamma} = e^{-\gamma(I-W)^{-1}} = \sum_{i=1}^{\infty}\frac{e^{-\gamma}\gamma^i}{i!} \left(\vecS W^i\right)	
\end{align*}
In \cite{fastmstp}, Banerjee and Lofgren extended \texttt{Bidirectional-PPR} to get bidirectional estimators for graph diffusions and other general Markov chain transition-probability estimation problems. These algorithms inherited similar performance guarantees to \texttt{Bidirectional-PPR} -- in particular, they had good expected running-time bounds for uniform-random choice of target node $t$. 
We now briefly discuss how we can modify \texttt{Undirected-BiPPR} to get an estimator for graph diffusions in undirected graphs with worst-case running-time bounds.

First, we observe that Lemma \ref{undirected_ppr_lemma} extends to all graph diffusions, as follows:
\begin{corollary} 
	\label{corr:undirected-mstp}
Let any undirected graph $G$ with random-walk matrix $W$, and any set of non-negative length weights $(\alpha_i)_{i=0}^\infty$ with $\sum \alpha_i=1$ be given. Define $f(W,\vecS)=\sum_{i=0}^{\infty}\alpha_i\left(\vecS W^{i}\right)$. Then for any node-pair $(s, t) \in V^2$, we have:
\[ f\left(W,\vecE_s\right) =  \frac{d_t}{d_s} f\left(W,\vecE_t\right). \]  
\end{corollary}
As before, the above result is stated for unweighted graphs, but it also extends to random-walks on weighted undirected graphs, if we define $d_i=\sum_jw_{ij}$.

Next, observe that for any graph diffusion $f(\cdot)$, the truncated sum $f^{\ell_{\max}}=\sum_{i=0}^{\ell_{\max}}\alpha_i\left(\PR_{\vecS}^TP^{i}\right)$ obeys: $||f-f^{\ell_{\max}}||_{\infty}\leq\sum_{\ell_{\max}+1}^{\infty}\alpha_k$
Thus a guarantee on an estimate for the truncated sum directly translates to a guarantee on the estimate for the diffusion.

The main idea in \cite{fastmstp} is to generalize the bidirectional estimators for PageRank to estimating \emph{multi-step transitions probabilities} (for short, MSTP).
Given a source node $s$, a target node $t$, and length $\ell\leq\ell_{\max}$, we define: 
$$p_s^{\ell}[t]=\PP[\mbox{Random-walk of length $\ell$ starting from $s$ terminates at $t$}]$$
Note from Corollary \ref{corr:undirected-mstp}, we have for any pair $(s,t)$ and any $\ell$, $p_s^{\ell}[t]d_s = p_t^{\ell}[s]d_t$.

Now in order to develop a bidirectional estimator for $p_s^{\ell}[t]$, we need to define a local-update step similar to \texttt{ApproximatePageRank}. For this, we can modify the \texttt{REVERSE-PUSH} algorithm from \cite{fastmstp}, as follows.

Similar to \texttt{ApproximatePageRank}, given a source node $s$ and maximum length $\ell_{\max}$, we associate with each length $\ell\leq\ell_{\max}$ an estimate vector $q_s^{\ell}$ and a residual vector $r_s^{\ell}$. 
%We initialize $r_s^{0}=\vecE_s$ and all other estimate vectors $\{q_s^{\ell}\}_{\ell=0}^{\ell_{\max}}$ and the residual vectors $\{r_s^{\ell}\}_{\ell=1}^{\ell_{\max}}$ as $0$.
These are updated via the following \texttt{ApproximateMSTP} algorithm:
\begin{algorithm}[!ht]
\caption{ApproximateMSTP$(G,s,\ell_{\max},\epr)$}
\label{alg:approxmstp}
\begin{algorithmic}[1]
\REQUIRE Graph $G$, source $s$, maximum steps $\ell_{\max}$, maximum residual $\epr$\\
\STATE Initialize: Estimate-vectors $q_s^k = \underline{0}\,,\, \fall k\in\{0,1,2,\ldots,\ell_{\max}\}$,\\
\hspace{1.4cm}  Residual-vectors $r_s^0 = \vecE_s$ and $r_s^k = \underline{0}\,,\, \fall k\in\{1,2,3,\ldots,\ell_{\max}\}$
\FOR{$i\in\{0,1,\ldots,\ell_{\max}\}$}
\WHILE{$\exists \, v\in \S\quad s.t.\quad r_t^i[v]/d_v>\epr$}
\FOR{$w\in\mathcal{N}(v)$}
\STATE $r_s^{i+1}[w] \pluseq r_s^{i}[v]/d_v$
\ENDFOR
\STATE $q_s^{i}[v] \pluseq r_s^{i}[v]$
\STATE $r_s^{i}[v] = 0$
\ENDWHILE
\ENDFOR
\RETURN $\{q_s^{\ell},r_s^{\ell}\}_{\ell=0}^{\ell_{\max}}$
\end{algorithmic}
\end{algorithm}    

The main observation now is that for any source $s$, target $t$, and length $\ell$, after executing the \texttt{ApproximateMSTP} algorithm, the vectors $\{q_s^{\ell},r_s^{\ell}\}_{\ell=0}^{\ell_{\max}}$ satisfy the following invariant (via a similar argument as in \cite{fastmstp}, Lemma $1$):
\begin{align*}
%\label{eq:pushinv}
p_{s}^{\ell}[t] 
%&= q_s^{\ell}[t] + \sum_{k=0}^{\ell}\langle r_s^{k}W^{\ell-k},\vecE_t\rangle 
= q_s^{\ell}[t] + \sum_{k=0}^{\ell}\sum_{v\in V} r_s^{k}[v]p_v^{\ell-k}[t]
&= q_s^{\ell}[t] + d_t\sum_{k=0}^{\ell}\sum_{v\in V} \frac{r_s^{k}[v]}{d_v}p_t^{\ell-k}[v]  	
\end{align*}	
As before, note now that the last term can be written as an expectation over random-walks originating from $t$. The remaining algorithm, accuracy analysis, and runtime analysis follow the same lines as those in Section \ref{sec:algo}.

\section{Lower Bound}
In \cite{fastppr}, the authors prove an average case lower bound for PPR-Estimation.  In particular they prove that there exists a family of undirected 3-regular graphs for which any algorithm that can distinguish between pairs $(s, t)$ with $\pi_s[t] > \delta$ and pairs $(s, t)$ with $\pi_s[t] < \frac{\delta}{2}$ (distinguishing correctly with constant probability 8/9), must access $\Omega(1/\sqrt{\delta})$ edges of the graph.  Since the algorithms in \cite{fastppr,bidir-search} solve this problem in time $O\left(\sqrt{\dbar / \delta}\right)$, where $\dbar$ is the average degree of the given graph, there remains a $\sqrt{d}$ gap between the lower bound and the best algorithm for the average case.  For the worst case (possibly parameterized by some property of the graph or target node), the authors are unaware of any lower bound stronger than this average case bound, and an interesting open question is to prove a lower bound for the worst case.

% We assume $\alpha = 1/100\log(1/\delta)$, and consider randomized algorithms for the following variant of Significant-PPR, which we denote as \signif$(\delta)$ -- for all pairs $(s,t)$:
% \begin{itemize}[nolistsep,noitemsep]
% \item If $\PR_s(t) > \delta$, output ACCEPT with probability $> 9/10$.
% \item If $\PR_s(t) < \frac{\delta}{2}$, output REJECT with probability $>9/10$.
% \end{itemize}
% We stress that the probability is over the random choices of the algorithm, \emph{not} over $s,t$. We now have the following lower bound:
% \begin{theorem} 
% \label{thm:lb} 
% Any algorithm for \signif$(\delta)$ must access $\Omega(1/\sqrt{\delta})$ edges of the graph.
% \end{theorem}

\section{Conclusion}

We have developed \texttt{Undirected-BiPPR}, a new bidirectional PPR-estimator for undirected graphs, which for any $(s,t)$ pair such that $\PR_s[t]>d_t/m$, returns an estimate with $\epsilon$ relative-error in worst-case running time of $O(\sqrt{m}/\epsilon)$. 
This thus extends the average-case running-time improvements achieved in \cite{fastppr,bidir-search} to worst-case bounds on undirected graphs, using the reversibility of random-walks on undirected graphs. 
Whether such worst-case running-time results extend to general graphs, or if PageRank computation is fundamentally easier on undirected graphs as opposed to directed graphs, remains an open question.

\section{Acknowledgments}
Research supported by the DARPA GRAPHS program via grant
FA9550-12-1-0411, and by NSF grant 1447697.  One author was supported by an NPSC fellowship.  Thanks to Aaron Sidford for a helpful discussion.

\bibliographystyle{abbrv}
\bibliography{PPR-refs}

\end{document}